\documentclass[11pt]{iopart}
\usepackage{bm}
\usepackage{hyperref}
\usepackage{amssymb}
\usepackage{amsopn}
\usepackage{setstack}

\newcommand{\fnl}{f_{\mathrm{NL}}}

\newcommand{\Planck}{M_{\mathrm{P}}}

\renewcommand{\H}{\mathcal{H}}
\DeclareMathOperator{\Gr}{Gr}
\newcommand{\F}{\mathcal{F}}
\newcommand{\G}{\mathcal{G}}
\newcommand{\src}{\mathcal{S}}
\newcommand{\symF}{\mathbb{F}}
\newcommand{\symG}{\mathbb{G}}

\renewcommand{\d}{\mathrm{d}}
\newcommand{\vect}[1]{\bm{\mathrm{{#1}}}}
\renewcommand{\e}[1]{\mathrm{e}^{{#1}}}
\newcommand{\im}{\mathrm{i}}

\renewcommand{\geq}{\geqslant}

\begin{document}
	\title{Non-gaussianity of inflationary field perturbations from
	the field equation}
	
	\author{David Seery$^1$, Karim A. Malik$^2$ and David H. Lyth$^3$}
	
	\address{$^1$ Centre for Theoretical Cosmology, \\
	Department of Applied Mathematics and Theoretical Physics, \\
	Wilberforce Road, Cambridge, CB3 0WA, United Kingdom \\ \vspace{3mm}
	$^2$ Astronomy Unit, School of Mathematical Sciences, \\
	Queen Mary University of London \\
	Mile End Road, London, E1 4NS, United Kingdom \\ \vspace{3mm}
	$^3$ Cosmology and Astroparticle Physics Group, \\
	Department of Physics, University of Lancaster, \\
	Lancaster, LA1 4YB, United Kingdom}

	\eads{\mailto{djs61@cam.ac.uk}}

	\begin{abstract}
		We calculate the tree-level bispectrum of the inflaton
		field perturbation directly from the field equations,
		and construct the corresponding $\fnl$ parameter. Our
		results agree with previous ones derived from the
		Lagrangian. We argue that quantum theory should only
		be used to calculate the correlators when they first
		become classical a few Hubble times after horizon
		exit, the classical evolution taking over thereafter.
	\vspace{3mm}
	\begin{flushleft}
	\end{flushleft}
	\end{abstract}
	\maketitle

	\section{Introduction}
	Recent advances in observational astronomy have allowed maps
	of the cosmic microwave background to be constructed in more
	detail than ever before \cite{Hinshaw:2006ia}.
	The availability of such maps, with good noise properties
	and controlled foreground subtraction, offers the exciting prospect
	of studying the earliest ages in the evolution of our universe in a
	relatively direct way \cite{Martin:2006rs,Kinney:2006qm}.
	
	In order to connect any theory of early universe physics with the cosmic
	microwave sky, one must make predictions for properties of the
	curvature perturbation, $\zeta$ \cite{Bardeen:1983qw},
	which sets the initial
	condition for those fluctuations in the matter and radiation densities
	which we can now observe as temperature anisotropies. To carry out this
	programme effectively, one needs both a model for the relevant physics
	and an efficient calculational tool with which to make predictions.
	
	One of the most attractive models which has been proposed to
	describe the evolution of the early universe is \emph{inflation}
	(for a review, see Ref. \cite{Liddle:2000cg}). In an inflationary scenario
	the universe is supposed to have undergone a phase of
	accelerated expansion in the very distant past. During this phase
	each light scalar field acquires a fluctuation, which is close to
	scale invariance when the rate at which the universe expands is almost
	constant. These fields each contribute a proportion of
	the total energy density of
	the universe, and the relative importance of these contributions
	is sufficient to determine how $\zeta$ is
	composed of the separate fluctuations in each light scalar
	\cite{Starobinsky:1986fx,Sasaki:1995aw,Lyth:2004gb,
	Langlois:2006vv,Rigopoulos:2004gr}.
	The fluctuations themselves can be computed, for example,
	by using the methods of quantum field theory.
	Inflation thus provides a framework within which the properties of
	$\zeta$ can readily be calculated, and for a large class of models
	these predictions are in extremely good agreement with the
	observational data \cite{Peiris:2003ff}.
	
	Which properties of $\zeta$ should we compute?
	Any measurement of the CMB anisotropy
	can be framed in terms of the $n$-point expectation values
	of the curvature perturbation, and
	current data are primarily sensitive to the simplest such
	expectation value, which is given by the two-point function
	or power spectrum.
	Although this observable has sufficed to obtain a good deal
	of precise information about the very early universe, there are limits
	to how much we can learn from it.
	Therefore, in order to discriminate effectively
	between the different models of inflation, it will be necessary to
	broaden our observational methods so that
	higher $n$-point functions of $\zeta$ become experimentally
	accessible, and at the same time develop theoretical predictions for these
	functions in all relevant models. On the experimental side this process
	has been underway since the \emph{Cosmic Background Observer} (COBE)
	satellite made the first all-sky map of the CMB in the mid 1990s
	\cite{Bennett:1996ce,Ferreira:1998kt}.
	Such efforts have borne fruit with the availability of more
	sensitive measurements taken by the NASA
	\emph{Wilkinson Microwave Anisotropy Probe} with high angular resolution
	\cite{Komatsu:2003fd,Peiris:2003ff},
	and will be improved still further by ESA's \emph{Planck}
	satellite, due for launch in mid-2008.
	The first year WMAP data release placed significant constraints on
	the level of non-gaussianity which could be present in the CMB
	(often quantified in terms of a so-called ``non-linearity parameter,''
	$\fnl$ \cite{Komatsu:2001rj}),
	and this quantitative estimate has subsequently been refined
	\cite{Creminelli:2006rz,Yadav:2007ny},
	leading to a recent high-confidence exclusion of
	$\fnl = 0$ \cite{Yadav:2007yy}.
	In the long-term, it may be possible to obtain even better estimates
	from observations, perhaps based on maps of the 21-cm emission
	of neutral hydrogen \cite{Loeb:2003ya,Cooray:2006km}.
	On the theoretical side, predictions for the non-gaussianity generated
	in a large collection of relevant models have become available
	over the last several years
	\cite{ArkaniHamed:2003uz,Alishahiha:2004eh,Rigopoulos:2004gr,
	Rigopoulos:2005xx,Rigopoulos:2005us,Lyth:2005fi,Lyth:2005qj,
	Zaballa:2006pv,Alabidi:2005qi,Seery:2005gb,Kim:2006te,Lyth:2004gb,
	Malik:2006pm,Sasaki:2006kq,Seery:2005wm},
	following Maldacena's successful calculation of
	the bispectrum produced in single-field, slow-roll inflation
	\cite{Maldacena:2002vr}.
	
	These developments mean that
	non-gaussianity is now a standard cosmological observable, comparable
	to the spectral index $n$ of the scalar power spectrum, and is a
	powerful discriminant between competing models. It is therefore
	extremely desirable to have at our disposal a simple calculational
	scheme---analogous to the familiar formula
	which allows the scalar spectral index to be estimated
	\cite{Liddle:2000cg}---which
	can be used to predict the non-gaussianity generated
	in any model of our choice.

	Most computations of the primordial non-gaussianity have been
	carried out using the tools of quantum field theory, and have been
	explicitly framed
	in the context of the Lagrangian formalism, which leads to a method of
	computation closely related to many calculations in particle physics
	\cite{Maldacena:2002vr,Seery:2005gb,Chen:2006nt}.
	For example,
	when expanded perturbatively, the Lagrangian formalism naturally gives
	rise to a variant of the Feynman diagrams which are familiar from the
	calculation of scattering amplitudes. There is no doubt that
	this is a useful development,
	which immediately permits methods and intuition developed in the
	context of quantum field theory to be imported into cosmology.
	Nevertheless,
	there is considerable merit in exploring alternative calculational
	strategies, either to enlarge the class of theories in which the
	non-gaussianity can be computed, or to take advantage of technical
	simplifications in the computation.

	In the present paper we return to the more familiar
	method of employing the field equations, which does not invoke a Lagrangian. 
	This is how the tree-level 
	spectrum $(H/2\pi)^2$ of a scalar field in the de Sitter spacetime
	was first calculated \cite{Birrell:1982ix}, and also how the tree-level
	spectrum of the curvature
	perturbation was first calculated using the Mukhanov--Sasaki field equation
	\cite{Mukhanov:1985rz,Sasaki:1986hm}.%
		\footnote
		{``Tree-level'' here means leading order in the perturbation, which indeed 
		corresponds to the tree-level Feynman graph \cite{Weinberg:2005vy,Musso:2006pt}.}
	More recently it has been shown how the method allows one
	to calculate all expectation values
	of field perturbation to all orders in de Sitter space
	\cite{Musso:2006pt}. Here we use the method to calculate the tree-level
	bispectrum of the inflaton field perturbation, to second order in the field
	perturbation and including the associated metric perturbation.

	The method of field equations 
	permits us to make contact
	with the cosmological literature, which has traditionally approached
	perturbation theory from the standpoint of the Einstein equations.
	A second advantage is that it allows an immediate calculation of
	the non-gaussianity generated in any model which can be cast in the
	form of an effective Klein--Gordon equation, whether or not
	the model is descended from an effective action principle.
	We do not invoke the Einstein--Hilbert action, only the Einstein field equation
	together with the inflaton field equation which is obtained from the flat-space
	Klein--Gordon equation by the replacement $\eta_{\mu\nu}\to g_{\mu\nu}$
	(which is the content of the equivalence principle).
	We are able to give a simple, essentially ``mechanical'' process for
	computing the bispectrum in such a model.
	
	The outline of this paper is as follows.
	In \S\ref{sec:construct}
	we construct an explicit solution for the Heisenberg picture field
	directly from the equations of motion.
	In \S\ref{sec:correlator} we use this solution to construct the
	three-point correlator of the field fluctuations. This is the essential
	ingredient in a calculation of the bispectum of $\zeta$, since it serves
	as the required initial condition for the well-known $\delta N$
	formula.
	In \S\ref{sec:fnl} we write down an explicit formula for the
	non-linearity parameter $\fnl$ before giving a brief summary
	of the calculation in \S\ref{sec:conclude}, which concludes with a
	discussion.
	
	Throughout this paper, we work in natural units where the
	Planck mass is set equal to unity, $\Planck \equiv (8\pi G)^{-1/2} = 1$.
	The background metric is taken to be of Robertson--Walker form,
	\begin{equation}
		\d s^2 = - \d t^2 + a(t)^2 \delta_{ij} \, \d x^i \, \d x^j ,
		\label{eq:metric}
	\end{equation}
	where the scale factor $a(t)$ obeys the usual Friedmann constraint
	$3 H^2 = \rho$, with $\rho$ being the energy density. 
	Throughout, $H \equiv \dot{a}/a$ is the Hubble parameter
	and an overdot denotes a derivative with respect to $t$. It is sometimes
	more convenient to work in terms of a conformal time variable,
	$\eta$, which is defined by
	$\eta \equiv\int_t^\infty \d t'/a(t')$. When $\eta$ is used, it is useful to
	define the conformal Hubble rate $\H$, which satisfies
	$\H \equiv a'/a = aH$.

	\section{Constructing the Heisenberg-picture field}
	\label{sec:construct}
	
	For simplicity we will assume
	there is only a single scalar degree of freedom, although our method
	generalizes easily to the case of many scalar fields.
	The perturbed metric can be written
	\begin{equation}
		\d s^2 = - N^2 \, \d t^2 +
			h_{ij} (\d x^i + N^i \, \d t) (\d x^j + N^j \, \d t) .
		\label{eq:adm-metric}
	\end{equation}
	We work in the uniform curvature gauge which corresponds to setting
	$h_{ij}$ equal to its unperturbed value
	[given by Eq.~\eref{eq:metric}], where we explicitly assume
	that all tensor modes are absent.%
	 	\footnote{A background of gravitational waves would not contribute to
		the bispectrum to leading order in the perturbations; its contribution
		would therefore be strictly subleading. For this reason, the truncation
		to zero tensor modes should be acceptable for the purpose of computing
		$\fnl$.}
	The quantities $N$ and $N^i$
	are, respectively, the lapse function and the shift vector
	\cite{Arnowitt:1962hi}. These are unambiguously
	determined by the Einstein constraint equations once a gauge has been
	chosen.
	
	Let $\phi$ be a canonically normalized field, whose background value
	sources the metric~\eref{eq:metric}. We assume that $\phi$
	is driven by a potential $V$, which supports inflation
	in some region of field space.	
	One then allows $\phi$ to be
	perturbed by a small amount $\delta\phi$, which obeys an
	effective Klein--Gordon equation. Perturbation theory
	is an expansion in powers of $\delta\phi$.
	
	In the cosmological literature a further decomposition is sometimes
	made (see, for example, Ref. \cite{Malik:2003mv}),
	by separating some part of $\delta\phi$ which can be called
	$\delta\phi_1$. One then writes all quantities in the perturbation theory,
	including $\delta\phi$, as an expansion in powers of $\delta\phi_1$,
	\begin{equation}
		\delta\phi \equiv \delta\phi_1 + \frac{1}{2} \delta\phi_2 +
		\cdots + \frac{1}{n!} \delta\phi_n + \cdots .
	\end{equation}
	It is necessary to impose some arbitrary auxiliary condition to determine
	$\delta\phi_1$, which is usually done implicitly by demanding that
	$\delta\phi_1$ obey a linear equation of motion. It follows that
	$\delta\phi_1$ exhibits precisely gaussian statistics. This formulation
	of perturbation theory is most useful when $\delta\phi$
	becomes increasingly gaussian at early times, which implies that the
	$\delta\phi_n$ for $n \geq 2$ vanish at past infinity. These two
	pictures of perturbation theory have complementary merits, and can be used
	interchangeably.
	
	We will work in the second picture and choose $\delta\phi_1$ to be
	that part of $\delta\phi$ which
	obeys precisely gaussian statistics, as discussed above,
	and formulate perturbation theory as an expansion in powers of
	$\delta\phi_1$. As we have described this implies that
	$\delta\phi_1$ should be chosen to obey a linear
	equation of motion, so that each Fourier mode evolves according
	to the equation
	\begin{equation}
		\delta\phi_1'' + 2 \H \delta\phi_1' + k^2 \delta\phi_1 = 0 ,
		\label{eq:free-KG}
	\end{equation}
	where a prime $'$ denotes a derivative with respect to $\eta$.
	Eventually we are aiming to calculate only to leading order in the
	slow-roll expansion and therefore higher-order slow-roll
	contributions in Eq.~\eref{eq:free-KG} have been omitted.
	It is useful to introduce the notation $\theta_k$ for solutions
	of~\eref{eq:free-KG}, so that
	$\delta\phi_1(\vect{k},\eta) \equiv \theta_k(\eta)$.
	
	So far, this theory is entirely classical.
	Quantization of $\delta\phi_1$ is straightforward, and proceeds by the
	canonical method (see, eg., Ref. \cite{Peskin:1995ev}).
	The quantum field corresponding to a conjugate pair of solutions
	$\{ \theta_k, \bar{\theta}_k \}$ is
	\begin{equation}
		\delta\hat{\phi}_1(\vect{x},\eta) = \int \frac{\d^3 k}{(2\pi)^3}
		\e{\im \vect{k}\cdot\vect{x}}
		\Big\{
			a^\ast_{\vect{k}} \theta_k(\eta) +
			a_{-\vect{k}} \bar{\theta}_k(\eta)
		\Big\} \,,
		\label{eq:quantum-phi}
	\end{equation}
	where the normalization of the $\{ \theta_k, \bar{\theta}_k \}$
	has been adjusted so that the canonical commutation relation
	for $\delta\hat{\phi}$ and its conjugate momentum is obtained when
	$\{ a_{\vect{k}}, a^\ast_{\vect{k}} \}$ obey the usual
	creation--annihilation algebra,
	\begin{equation}
		[ a_{\vect{k}}, a^\ast_{\vect{k}'} ] = (2\pi)^3
			\delta(\vect{k}-\vect{k}') .
		\label{eq:algebra}
	\end{equation}
	
	On the other hand, $\delta\phi_2$ obeys a non-linear equation
	with quadratic source terms.
	In this paper we are going to use the slow-roll approximation
	to control these sources.
	As usual we take slow-roll
	to be defined by the following set of conditions:
	\begin{eqnarray}
		\epsilon & \equiv & - \frac{\dot{H}}{H^2} \simeq
		\frac{1}{2} \left( \frac{V'}{V} \right)^2 \simeq
		\frac{1}{2} \frac{\dot{\phi}^2}{H^2} \ll 1
		\label{first} , \\
		\eta & \equiv & \frac{V''}{V} \simeq
		- \frac{\ddot{\phi}}{H \dot{\phi}} + \epsilon
		\ll 1 \label{second} .
	\end{eqnarray}
	(In particular, one should take
	care that $\eta$ is not confused with the conformal time.)
	The non-linear equation which describes $\delta\phi_2$
	can now be written in the form
	\begin{equation}
		\delta\phi_2'' + 2 \H \delta\phi_2' + k^2 \delta\phi_2 =
		- a^2 V''' \delta\phi_1^2 +
		F_2(\delta\phi_1) + G_2(\delta\phi'_1),
		\label{eq:interacting-KG}
	\end{equation}
	where $V'''$ is the third derivative of the potential for $\phi$, and
	$F_2$ and $G_2$ are source terms, which are respectively
	quadratic in $\delta\phi_1$ and $\delta\phi'_1$.
	It has recently been shown that
	\cite{Malik:2006ir,Malik:2007nd}
	\begin{equation}
		\fl
		F_2 \equiv - \sqrt{2\epsilon} \left\{
			\frac{1}{2} \partial \delta\phi_1 \partial \delta\phi_1 -
			2 \delta\phi_1 \partial^2 \delta\phi_1 -
			\partial^{-2} \left[\partial^2\delta\phi_1\partial^2 \delta\phi_1 +
				\partial \delta\phi_1 \partial \partial^2 \delta\phi_1 \right]
			\right\} ,
		\label{eq:ftwo}
	\end{equation}
	and
	\begin{equation}
		\fl
		G_2 \equiv - \sqrt{2\epsilon} \left\{
			\frac{1}{2} \delta\phi'_1 \delta\phi'_1 + 2 \partial \delta\phi'_1
			\partial \partial^{-2} \delta\phi'_1 -
			\partial^{-2} \left[ \delta\phi'_1 \partial^2 \delta\phi'_1 +
				\partial \delta\phi'_1 \partial \delta\phi'_1 \right]
			\right\} ,
		\label{eq:gtwo}
	\end{equation}
	where we have dropped terms that are suppressed by factors of $\epsilon$ and
	$\eta$.
	It is useful to separate the $a^2 V'''$ term from the remainder,
	because its different
	dependence on the scale factor endows it with a distinctive time
	dependence which we shall see below leads to a logarithmic divergence
	at late times.

	Among the small terms---subleading in powers of $\epsilon$ and
	$\eta$---which have been dropped in
	Eqs.~\eref{eq:interacting-KG}--\eref{eq:gtwo}
	are source terms proportional to the cross-product
	$\delta\phi_1 \delta\phi'_1$.
	Such terms also lead to logarithmic divergences which have the rough
	form $\sim \epsilon_\ast^{3/2} N$, where for any mode of wavelength
	$k$ the quantity $N \equiv \ln |k\eta|$
	measures by how many e-foldings it is outside the horizon at some
	time of observation, $\eta$, and a subscript `$\ast$' denotes
	evaluation at horizon crossing.
	Since we are assuming that slow-roll applies at that time
	these terms are small when
	$N \approx 0$, but they grow in magnitude with $N$ and when
	$N \sim 1/\epsilon_\ast$ they can no longer be ignored.
	Indeed, such terms form part of a tower of divergences which have the
	rough form $\sim \epsilon_\ast^{m+1/2} N^m$ for $m \geq 1$.
	It follows that the slow-roll approximation will be close to breaking
	down if we wish to evaluate expectation values at the end of inflation,
	where the scales of interest are $N$ e-foldings outside the horizon
	(with values of order $N \sim 60$ being typical), but in many
	theories $\epsilon_\ast \sim 1/N$.
	This would lead to a nonsensical perturbation theory in powers of unity.

	This argument shows that we cannot use an expansion in powers of slow-roll
	parameters to predict expectation values at the end of inflation.
	One should instead think of these divergences as terms in a
	Taylor series expansion around the time of horizon exit, where $N = 0$
	\cite{Gong:2001he}.
	In principle they could be resummed by the method of the
	renormalization group or an equivalent technique, after which they
	would merely correspond to the classical time evolution
	\cite{Zaldarriaga:2003my,Matarrese:2007wc,Seery:2007wf,Bartolo:2007ti}.
	However, there is no need to perform such a complicated resummation.
	Instead, we believe it is most accurate to evaluate all expectation
	values almost immediately after horizon exit, where slow-roll is an
	excellent approximation and all strictly
	positive powers of $N$ are entirely negligible.
	One must then use some other method, which does not rely on an expansion
	in terms of e-foldings since horizon exit, to evolve these
	expectation values until the desired time of evaluation.
	This is equivalent to the argument of the renormalization group, but
	makes use of the known simplification that evolution outside the horizon
	is simply classical.
	For the curvature perturbation the
	$\delta N$ formula is very convenient
	\cite{Starobinsky:1986fx,Birrell:1982ix,Sasaki:1995aw,Zaldarriaga:2003my,Lyth:2004gb}.
	More generally one can use the separate universe approach
	\cite{Starobinsky:1986fxa,Sasaki:1995aw,Lyth:2004gb,Lyth:2005fi,Lyth:2006gd},
	or an equivalent gradient expansion
	\cite{Langlois:2006vv,Rigopoulos:2004gr}. It may also be possible
	to use some formulations of conventional perturbation theory, provided
	they avoid the appearance of powers of $N$
	\cite{Malik:2006ir}. The crucial point is that
	if the correlation functions are evaluated immediately after
	horizon exit, it is only necessary to
	compute the leading term in the Taylor series, which will be the constant
	term in any model giving rise to a Klein--Gordon equation of the form
	\eref{eq:interacting-KG}. This term is given by the lowest-order
	slow-roll approximation.

	In theories which are
	more general than Einstein gravity coupled to a scalar field,
	one might encounter examples where the constant term in the Taylor expansion
	is absent, or where the linear term has a comparable magnitude.
	An example of such a model is the case of non-local inflation studied by
	Barnaby \& Cline \cite{Barnaby:2007yb}, and we shall see below that it can
	occur even in single-field, slow-roll inflation if the third derivative of
	the potential is exceptionally large in comparison with the first and second.
	In such cases one can compute enough terms in the Taylor series to obtain a
	satisfactory approximation before using the separate universe picture
	or some equivalent technique to obtain the superhorizon evolution.
	If any $\delta\phi_1 \delta\phi_1'$ terms participate in this process, then
	they can be computed using a small modification of the argument outlined
	below.
	
	In the classical theory, Eq.~\eref{eq:interacting-KG} can be solved by
	the use of a Green's function. The correct choice is the so-called
	causal or retarded Green's function, which in momentum space satisfies
	\begin{equation}
		\Gr_k(\eta,\tau) = \im a(\tau)^2 \times \left\{
			\begin{array}{l@{\hspace{5mm}}l}
				0	& \eta < \tau \\
				\theta_k(\tau) \bar{\theta}_k(\eta) -
				\bar{\theta}_k(\tau) \theta_k(\eta) & \eta > \tau
			\end{array} \right. ,
	\end{equation}
	where $\{ \theta_k, \bar{\theta}_k \}$ are a complex conjugate pair of
	solutions to the non-interacting Klein--Gordon equation~\eref{eq:free-KG}
	for comoving wavenumber $k$, normalized as in
	Eqs.~\eref{eq:quantum-phi}--\eref{eq:algebra},
	and a bar denotes complex conjugation.
	To use $\Gr_k$ to solve for $\delta\phi_2$, one must translate the
	source terms $F_2$ and $G_2$ into Fourier modes. The behaviour of each
	interacting mode can then be computed, and the result reassembled,
	giving
	\begin{equation}
		\fl
		\delta\phi_2(\eta,\vect{x}) =
			\int \frac{\d^3 q}{(2\pi)^3} \e{\im \vect{q}\cdot\vect{x}}
			\left\{
				\int_{-\infty}^{\eta} \d \tau
				\int \frac{\d^3 k_1 \, \d^3 k_2}{(2\pi)^6}
				\Gr_q(\eta,\tau)
				\delta(\vect{q} - \vect{k}_1 - \vect{k}_2) \src_2
			\right\} ,
	\end{equation}
	where $\src_2$ is defined by
	\begin{equation}
		\fl
		\src_2 \equiv \left[ - a^2(\tau) V''' + \F_2(\vect{k}_1,\vect{k}_2)
			\right]
			\delta\phi_1(\vect{k}_1,\tau) \delta\phi_1(\vect{k}_2,\tau) +
			\G_2(\vect{k}_1,\vect{k}_2)
			\frac{\d\delta\phi_1(\vect{k}_1,\tau)}{\d\tau}
			\frac{\d\delta\phi_1(\vect{k}_2,\tau)}{\d\tau} ,
	\end{equation}
	and the momentum space source functions
	$\F_2$ and $\G_2$ satisfy
	\begin{eqnarray}
		\F_2 & \equiv & - \frac{\phi'}{\H} \left(
			- \frac{1}{2} \vect{k}_1 \cdot \vect{k}_2 + 2 k_2^2 +
			\frac{1}{(\vect{k}_1 + \vect{k}_2)^2} \left[
				k_1^2 k_2^2 + k_2^2 \vect{k}_1 \cdot \vect{k}_2 \right]
			\right) \\
		\G_2 & \equiv & - \frac{\phi'}{\H} \left(
			\frac{1}{2} + \frac{2}{k_2^2} \vect{k}_1 \cdot \vect{k}_2 -
			\frac{1}{(\vect{k}_1 + \vect{k}_2)^2} \left[
				k_2^2 + \vect{k}_1 \cdot \vect{k}_2 \right]
			\right) .
	\end{eqnarray}
	
	In the quantum theory the same construction applies, with the
	understanding that we are to substitute the quantum field
	$\delta\hat{\phi}_1(\vect{q}) \equiv
	a^\ast_{\vect{q}} \theta_q(\eta) + a_{-\vect{q}}
	\bar{\theta}_q(\eta)$ for the non-interacting solution
	$\delta\phi_1(\vect{q}) \equiv \theta_q$ in the source term
	$\src_2$. Thus, once the quantization of $\delta\phi_1$ has been
	determined, this is sufficient to fix the quantization of
	the $\delta\phi_n$ for $n \geq 2$.
	
	\section{The three-point correlator}
	\label{sec:correlator}
	
	The final step is to use our solution for the interacting Heisenberg
	field to compute the three-point expectation value,
	$\langle \delta\phi(\vect{k}_1) \delta\phi(\vect{k}_2)
	\delta\phi(\vect{k}_3) \rangle$.
	Since $\delta\phi_1$ obeys precisely gaussian statistics, it is
	forbidden to have a non-trivial three-point correlator.
	The leading contribution therefore comes from a correlation between the
	second-order part of one of the fields with the first-order
	part of the remaining two, so that the correlation is schematically
	of the form
	$\langle \delta\phi_1 \delta\phi_1 \frac{1}{2} \delta\phi_2 \rangle \sim
	\frac{1}{2}
	\langle \delta\phi_1 \delta\phi_1 \delta\phi_1 \ast \delta\phi_1 \rangle$,
	where `$\ast$' denotes a convolution.
	The resulting four-$\delta\phi_1$ expectation can be evaluated by use
	of Wick's theorem, together with the Wightman function for
	$\delta\phi_1$,
	\begin{equation}
		\langle \delta\phi_1(\vect{k},\eta) \delta\phi_1(\vect{k}',\eta')
		\rangle = (2\pi)^3\delta(\vect{k} + \vect{k}') \bar{\theta}_{k}(\eta)
		\theta_{k}(\eta') .
		\label{eq:wightman}
	\end{equation}
	Notice that there is no time ordering in Eq.~\eref{eq:wightman}.
	
	\subsection{The $V'''$ term}
	Consider first the $V'''$ term, leading to
	a logarithmic divergence as described above.
	The three-point expectation value sourced by a term of this type
	was first calculated by Falk, Rangarajan and Srednicki
	\cite{Falk:1992sf}, and was later reconsidered by
	Zaldarriaga, who calculated it in the
	context of the inhomogeneous decay rate model \cite{Zaldarriaga:2003my}.
	
	This term gives a contribution to the three-point function which can
	be written
	\begin{eqnarray}
		\fl\nonumber
		\langle \delta\phi(\vect{k}_1) \delta\phi(\vect{k}_2)
		\delta\phi(\vect{k}_3) \rangle & \supseteq &
		- \im (2\pi)^3 \delta(\vect{k}_1 + \vect{k}_2 + \vect{k}_3)
		\int_{-\infty}^{\eta} \d\tau \; a(\tau)^4 V''' \times \\
		\nonumber & & \hspace{-2.3cm}
		\Bigg\{ \left[
			\theta_{k_3}(\tau)\bar{\theta}_{k_3}(\eta) -
			\bar{\theta}_{k_3}(\tau)\theta_{k_3}(\eta) \right]
		\bar{\theta}_{k_1}(\eta) \bar{\theta}_{k_2}(\eta)
		\theta_{k_1}(\tau) \theta_{k_2}(\tau) + \mbox{} \\
		\nonumber & & \hspace{-2cm}
		\left[
			\theta_{k_2}(\tau)\bar{\theta}_{k_2}(\eta) -
			\bar{\theta}_{k_2}(\tau)\theta_{k_2}(\eta) \right]
		\bar{\theta}_{k_1}(\eta) \theta_{k_3}(\eta)
		\theta_{k_1}(\tau) \bar{\theta}_{k_3}(\tau) + \mbox{} \\
		& & \hspace{-2cm}
		\left[
			\theta_{k_1}(\tau)\bar{\theta}_{k_1}(\eta) -
			\bar{\theta}_{k_1}(\tau)\theta_{k_1}(\eta) \right]
		\theta_{k_2}(\eta) \theta_{k_3}(\eta)
		\bar{\theta}_{k_2}(\tau) \bar{\theta}_{k_3}(\tau) \Bigg\} .
		\label{eq:xi-term}
	\end{eqnarray}
	It is easy to verify that four of the six terms in \eref{eq:xi-term}
	cancel against each other, leaving the sum of a quantity and
	its complex conjugate,
	\begin{eqnarray}
		\fl\nonumber
		\langle \delta\phi(\vect{k}_1) \delta\phi(\vect{k}_2)
		\delta\phi(\vect{k}_3) \rangle & \supseteq &
		- \im (2\pi)^3 \delta(\vect{k}_1 + \vect{k}_2 + \vect{k}_3)
		\int_{-\infty}^{\eta} \d\tau \; a(\tau)^4 V'''
		\prod_i \theta_{k_i}(\tau) \bar{\theta}_{k_i}(\eta) \\ & & \quad
		+ \mbox{complex conjugate} .
		\label{eq:xi}
	\end{eqnarray}
	where $i \in \{ 1, 2, 3 \}$. Such cancellation
	always occurs in the Lagrangian version of the in--in formalism, leading
	to expectation values which are manifestly real. In the present
	formalism any cancellation is contingent on the possibility of
	factoring the $\theta$, $\bar{\theta}$ terms. As we shall see below
	this does not always occur, leading to calculations which are somewhat
	longer than the Lagrangian equivalent and to the loss of manifest
	reality. At the end of the calculation, of course, the two methods must
	produce equivalent results.
	
	The integral in~\eref{eq:xi} is badly divergent, but almost all the
	divergent terms are imaginary and cancel out of the final result.
	This is easiest to see after integration by parts, which isolates
	the divergent pieces. One finds%
		\footnote{Note that this result differs from that given by
		Zaldarriaga by a sign and the replacement
		$\gamma \mapsto \gamma + 1/3$. We thank Filippo Vernizzi
		for pointing this out to us.}
	\begin{eqnarray}
		\fl\nonumber
		\langle \delta\phi(\vect{k}_1) \delta\phi(\vect{k}_2)
		\delta\phi(\vect{k}_3) \rangle
		& \supseteq & (2\pi)^3 \delta(\vect{k}_1 + \vect{k}_2 + \vect{k}_3)
		\frac{H_\ast^2 V'''_\ast}{4 \prod_i k_i^3} \times \\ & &
		\quad \left(
			- \frac{4}{9} k_t^3 + k_t \sum_{i < j} k_i k_j +
			\frac{1}{3} \Big\{ \frac{1}{3} + \gamma +
				\ln | k_t \eta_\ast | \Big\} \sum_i k_i^3
		\right) \label{eq:xi-expectation},
	\end{eqnarray}
	where $i \in \{ 1, 2, 3 \}$, $k_t = \sum_i k_i$,
	and for the purposes of obtaining an analytic solution
	we have assumed that the $k_i$ have
	some approximately equal magnitude, which defines a more or less
	unique time of horizon exit $\eta_\ast$. A subscript `$\ast$' denotes
	evaluation at this time, and $\gamma \approx 0.577216$
	is Euler's constant.
	
	\subsection{Zero-derivative terms}
	Now let us focus on those quadratic
	source terms in Eq.~\eref{eq:interacting-KG} which enter with no
	time derivatives, described by the function $F_2$.
	Such terms make a contribution to the three-point expectation value
	of the form
	\begin{eqnarray}
		\fl\nonumber
		\langle \delta\phi(\vect{k}_1) \delta\phi(\vect{k}_2)
		\delta\phi(\vect{k}_3) \rangle & \supseteq &
		\im (2\pi)^3 \delta(\vect{k}_1 + \vect{k}_2 + \vect{k}_3)
		\int_{-\infty}^{\eta_\ast} \d\tau \; a(\tau)^2 \times \\
		\nonumber & & \hspace{-2.3cm}
		\Bigg\{ \symF_2(k_1,k_2;k_3) \left[
			\theta_{k_3}(\tau)\bar{\theta}_{k_3}(\eta) -
			\bar{\theta}_{k_3}(\tau)\theta_{k_3}(\eta) \right]
		\bar{\theta}_{k_1}(\eta) \bar{\theta}_{k_2}(\eta)
		\theta_{k_1}(\tau) \theta_{k_2}(\tau) + \mbox{} \\
		\nonumber & & \hspace{-2cm}
		\symF_2(k_1,k_3;k_2) \left[
			\theta_{k_2}(\tau)\bar{\theta}_{k_2}(\eta) -
			\bar{\theta}_{k_2}(\tau)\theta_{k_2}(\eta) \right]
		\bar{\theta}_{k_1}(\eta) \theta_{k_3}(\eta)
		\theta_{k_1}(\tau) \bar{\theta}_{k_3}(\tau) + \mbox{} \\
		& & \hspace{-2cm}
		\symF_2(k_2,k_3;k_1) \left[
			\theta_{k_1}(\tau)\bar{\theta}_{k_1}(\eta) -
			\bar{\theta}_{k_1}(\tau)\theta_{k_1}(\eta) \right]
		\theta_{k_2}(\eta) \theta_{k_3}(\eta)
		\bar{\theta}_{k_2}(\tau) \bar{\theta}_{k_3}(\tau) \Bigg\} ,
		\label{eq:f-terms}
	\end{eqnarray}
	where the fields which participate in the expectation value are
	each evaluated at time $\eta_\ast$. The quantity
	$\symF_2(a,b;c)$ is to be obtained by symmetrizing
	$\F_2(\vect{a},\vect{b})$ between $\vect{a}$ and $\vect{b}$
	with weight unity. (This follows from the factor of $1/2$ which is used
	in the definition of $\delta\phi_2$; if this factor is absent, the
	the relative weighting of $\symF_2$ must be adjusted to match.)
	However, the result is subject to the constraint
	$\vect{a} + \vect{b} + \vect{c} = 0$ and therefore can be written in
	several equivalent forms. For this reason it is useful to
	fix the representation unambiguously, following Maldacena
	\cite{Maldacena:2002vr}, by using
	the relation between $\{ \vect{a}, \vect{b}, \vect{c} \}$
	to eliminate all dot products between distinct vectors, leaving a result
	which depends only on the magnitudes $\{ a, b, c \}$. We find
	\begin{equation}
		\symF_2(k_1,k_2;k_3) \equiv - \frac{\phi'}{\H} \left(
		\frac{3}{2} (k_1^2 + k_2^2) - \frac{(k_1^2 - k_2^2)^2}{4 k_3^2} -
		\frac{k_3^2}{4} \right) .
	\end{equation}
	
	If $\symF_2(a,b;c)$ were symmetric over permutations of
	$\{ a, b, c \}$, then as above the six terms
	in Eq.~\eref{eq:f-terms} would partially cancel among themselves.
	Owing to its construction, however,
	$\symF_2$  is in general symmetric only under interchange of $a$ and
	$b$ and fails to exhibit any symmetry under exchange of $c$ with
	$\{ a, b \}$. Therefore, it is not obvious that
	Eq.~\eref{eq:f-terms} is real. To go further and demonstrate equality with
	the Lagrangian formulation, it is necessary to compute each of the six
	integrals.
	
	Once this is done, we find that the zero-derivative contribution to the
	expectation value becomes
	\begin{eqnarray}
		\fl\nonumber
		\langle \delta\phi(\vect{k}_1) \delta\phi(\vect{k}_2)
		\delta\phi(\vect{k}_3) \rangle & \supseteq &
		(2\pi)^3 \delta(\vect{k}_1 + \vect{k}_2 + \vect{k}_3) \times \mbox{} \\
		& & \hspace{-1cm}
		\frac{H_\ast^4}{8 \prod_i k_i^3}
		\Big\{ f_1 \symF_2(k_1,k_2;k_3) +
			f_2 \symF_2(k_1,k_3;k_2) +
			f_3 \symF_2(k_2,k_3;k_1) \Big\} ,
		\label{eq:f-correlator}
	\end{eqnarray}
	where the momentum factors $f_i$ obey
	\begin{eqnarray}
		f_1 & \equiv & - \frac{2 k_3^3(k_1^2 + 4 k_1 k_2 + k_2^2 - k_3^2)}
								{(k_1 + k_2 - k_3)^2 k_t^2}\,, \label{eq:fa}\nonumber \\
		f_2 & \equiv & - \frac{2 k_2^3(k_1^2 - 4 k_1 k_3 + k_3^2 - k_2^2)}
								{(k_2^2 - (k_1 - k_3)^2)^2}\,, \label{eq:fb}\nonumber \\
		f_3 & \equiv & - \frac{2 k_1^3(k_2^2 + 4 k_2 k_3 + k_3^2 - k_1^2)}
								{(k_1 + k_2 - k_3)^2 k_t^2} \label{eq:fc} ,
	\end{eqnarray}
	and we have again assumed that the $k_i$ have an approximate common value, $k$.
	The expectation value is to be evaluated at
	a time just after horizon exit of the mode with wavenumber $k$.
	As above, $H_\ast$ denotes the value of the Hubble parameter at this time.
	Note that the $f_i$ do not vary from theory to theory; they are fixed
	by the structure of the interaction, and once calculated do not need
	to be calculated again. The different theories which may govern
	the interactions of $\delta\phi$ influence the
	three-point expectation only through the term $\symF_2(a,b;c)$. Similar
	remarks apply for the $n$-point correlation functions for all $n$.
	
	One might have expected that the $f_i$ should be related by
	cyclic permutations among the various $k_i$. In fact there are
	extra signs which are introduced, because the $f_i$ arise
	from integrating over combinations of the mode functions $\theta_{k_i}$
	and their conjugates $\bar{\theta}_{k_i}$. This explains the
	apparent lack of symmetry in Eq.~\eref{eq:fc}.
	
	\subsection{Two-derivative terms}
	We can carry out a similar calculation as in the previous section for the
	terms in the $\delta\phi_2$ equation which are quadratic in
	$\delta\phi_1'$, given in \eref{eq:gtwo}.
	One arrives at a contribution to the three-point
	expectation which can be written
	\begin{eqnarray}
		\fl\nonumber
		\langle \delta\phi(\vect{k}_1) \delta\phi(\vect{k}_2)
		\delta\phi(\vect{k}_3) \rangle & \supseteq &
		\im (2\pi)^3 \delta(\vect{k}_1 + \vect{k}_2 + \vect{k}_3)
		\int_{-\infty}^{\eta_\ast} \d\tau \; a(\tau)^2 \times \\
		\nonumber & & \hspace{-3.3cm}
		\Bigg\{ \symG_2(k_1,k_2;k_3) \left[
			\theta_{k_3}(\tau)\bar{\theta}_{k_3}(\eta) -
			\bar{\theta}_{k_3}(\tau)\theta_{k_3}(\eta) \right]
		\bar{\theta}_{k_1}(\eta) \bar{\theta}_{k_2}(\eta)
		\frac{\d\theta_{k_1}(\tau)}{\d\tau}
		\frac{\d\theta_{k_2}(\tau)}{\d\tau} + \mbox{} \\
		\nonumber & & \hspace{-3cm}
		\symG_2(k_1,k_3;k_2) \left[
			\theta_{k_2}(\tau)\bar{\theta}_{k_2}(\eta) -
			\bar{\theta}_{k_2}(\tau)\theta_{k_2}(\eta) \right]
		\bar{\theta}_{k_1}(\eta) \theta_{k_3}(\eta)
		\frac{\d\theta_{k_1}(\tau)}{\d\tau}
		\frac{\d\bar{\theta}_{k_3}(\tau)}{\d\tau} + \mbox{} \\
		& & \hspace{-3cm}
		\symG_2(k_2,k_3;k_1) \left[
			\theta_{k_1}(\tau)\bar{\theta}_{k_1}(\eta) -
			\bar{\theta}_{k_1}(\tau)\theta_{k_1}(\eta) \right]
		\theta_{k_2}(\eta) \theta_{k_3}(\eta)
		\frac{\d\bar{\theta}_{k_2}(\tau)}{\d\tau}
		\frac{\d\bar{\theta}_{k_3}(\tau)}{\d\tau} \Bigg\} ,
		\label{eq:g-terms}
	\end{eqnarray}
	where $\symG_2(a,b;c)$ is obtained from $\G_2(\vect{a},\vect{b})$
	by a rule analogous to that relating $\symF_2$ and $\F_2$:
	one symmetrizes over $\vect{a}$ and $\vect{b}$, and then uses the
	relation $\vect{a} + \vect{b} + \vect{c} = 0$ to eliminate dot
	products of distinct vectors. This gives
	\begin{equation}
		\symG_2(k_1,k_2;k_3) \equiv \frac{\phi'}{\H}
		\frac{(k_1^2 + k_2^2)(k_1^2 + k_2^2 - k_3^2)}{2 k_1^2 k_2^2}
		.
	\end{equation}
	
	In Eq.~\eref{eq:g-terms} there is no cancellation between terms
	even in the case that $\symG_2(a,b,c)$ is totally symmetric.
	However, we can follow a similar line of argument used in the case
	of the zero-derivative terms and find
	\begin{eqnarray}
		\fl\nonumber
		\langle \delta\phi(\vect{k}_1) \delta\phi(\vect{k}_2)
		\delta\phi(\vect{k}_3) \rangle & \supseteq &
		(2\pi)^3 \delta(\vect{k}_1 + \vect{k}_2 + \vect{k}_3) \times \mbox{} \\
		& & \hspace{-1cm}
		\frac{H_\ast^4}{8 \prod_i k_i^3}
		\Big\{ g_1 \symG_2(k_1,k_2;k_3) +
			g_2 \symG_2(k_1,k_3;k_2) +
			g_3 \symG_2(k_2,k_3;k_1) \Big\} .
		\label{eq:g-correlator}
	\end{eqnarray}
	The two-derivative momentum factors $g_i$ are defined by
	\begin{eqnarray}
		g_1 & \equiv & \frac{4 \prod_i k_i^2}
							{(k_1 + k_2 - k_3)^2 k_t^2}\,, \nonumber \\
		g_2 & \equiv & \frac{4 \prod_i k_i^2}
							{(k_2^2 - (k_1 - k_3)^2)^2}\,, \nonumber \\
		g_3 & \equiv & \frac{4 \prod_i k_i^2}
							{(k_1^2 - (k_2 + k_3)^2)^2}\,.
	\end{eqnarray}
	
	We can now assemble the zero- and two-derivative
	terms to obtain an overall contribution to
	the three-point expectation value which corresponds to
	\begin{eqnarray}
		\fl\nonumber
		\langle \delta\phi(\vect{k}_1) \delta\phi(\vect{k}_2)
		\delta\phi(\vect{k}_3) \rangle & \supseteq &
		(2\pi)^3 \delta(\vect{k}_1 + \vect{k}_2 + \vect{k}_3) \times \mbox{} \\
		& & \quad
		\frac{H_\ast^4}{8 \prod_i k_i^3} \frac{\dot{\phi}_\ast}{H_\ast}
		\left\{
			\frac{1}{2} \sum_i k_i^3 -
			\frac{4}{k_t} \sum_{i < j} k_i^2 k_j^2 -
			\frac{1}{2} \sum_{i \neq j} k_i k_j^2
		\right\} ,
	\end{eqnarray}
	in complete agreement with previous calculations using the action-based
	approach \cite{Seery:2005gb,Vernizzi:2006ve}.
	
	\section{The non-linearity parameter $\fnl$}
	\label{sec:fnl}
	The three-point correlator of $\delta\phi$
	is not itself the object of primary
	interest, since it is not directly observable in the cosmic microwave
	background. Instead, experiments conventionally set limits on a
	parameter $\fnl$ which is related to the three-point correlator of the
	curvature perturbation, $\zeta$.
	
	We define $\fnl$ by
	\begin{equation}
		\langle \zeta(\vect{k}_1) \zeta(\vect{k}_2) \zeta(\vect{k}_3) \rangle
		\equiv (2\pi)^3 \delta(\vect{k}_1 + \vect{k}_2 + \vect{k}_3)
		\frac{6}{5} \fnl \sum_{i < j} P_\zeta(k_i) P_\zeta(k_j) .
		\label{eq:fnl-def}
	\end{equation}
	With this choice of sign and prefactor, the parameter $\fnl$ corresponds
	to the one conventionally used in the analysis of CMB observations
	\cite{Komatsu:2003iq},
	although other sign conventions exist \cite{Maldacena:2002vr}.

	For a general theory one can use the $\delta N$ formula to show that 
	\begin{equation}
		\zeta(\vect{x}) \equiv
		\delta N = \frac{\partial N}{\partial \phi_\ast}
		\delta\phi_\ast(\vect{x}) +
		\frac{1}{2} \frac{\partial^2 N}{\partial \phi_\ast^2}
		\delta\phi_\ast(\vect{x})^2 + \cdots ,
	\end{equation}
	which holds in coordinate space and
	is easily generalized to the case of multiple fields. In this
	equation, the $\delta\phi$ are evaluated on a flat hypersurface at
	time $\eta_\ast$, and $N$ is a function of \emph{both} $\eta_\ast$ and
	the time of observation.
	To second order in $\delta\phi$, $\zeta$ can be calculated in terms of the
	so-called ``separate universe approach,'' and takes the form%
	\footnote{In writing this expression, we are assuming $\partial V/\partial \phi > 0$
	which can be accommodated by reversing the sign of $\phi$ if necessary.}
	\begin{equation}
		\zeta(\vect{k}) = \frac{1}{\sqrt{2 \epsilon_\ast}}
		\delta\phi_\ast(\vect{k})
		+ \frac{1}{2} \left( 1 - \frac{\eta_\ast}{2\epsilon_\ast} \right)
		\int \frac{\d^3 q}{(2\pi)^3} \;
		\delta\phi_\ast(\vect{k}_1 - \vect{q}) \delta\phi_\ast(\vect{q}) +
		\cdots ,
		\label{eq:zeta}
	\end{equation}
	where `$\cdots$' denotes terms of higher order in $\delta\phi_\ast$
	which have been omitted. In this expression, $\ast$ denotes evaluation
	at some time $t$ after the mode with wavenumber $k$ has left the
	horizon, after which $\zeta(\vect{k})$ maintains a constant value.

	To complete the calculation of $\fnl$ we will use
	two more slow-roll parameters:
	\begin{eqnarray}
		\xi^2 & \equiv & \frac{V'V'''}{V^2} \\
		\sigma^3 & \equiv &
		\frac{(V')^2(\d^4 V/\d\phi^4)}{V^3} \label{sigdef} ,
	\end{eqnarray}
	and the slow-roll relations
	\begin{equation}
		- \frac{\d(\ln \epsilon_\ast)}{\d N_\ast}
		\simeq 4\epsilon_\ast -2\eta_\ast, \label{depsilon}
	\end{equation}
	\begin{equation}
		- \frac{\d\eta_\ast}{\d N_\ast} \simeq
		2\epsilon_\ast\eta_\ast - \xi_\ast^2, \label{deta}
	\end{equation}
	and
	\begin{equation}
		-\frac{\d\xi_\ast^2}{\d N_\ast} \simeq
		4\epsilon_\ast\xi_\ast^2 - \eta_\ast\xi_\ast^2 - \sigma_\ast^3
		\label{dxi} .
	\end{equation}
	The slow-roll approximation, which we defined through
	Eqs.~\eref{first}--\eref{second}, does not in itself place
	conditions on $\xi^2$ and $\sigma^3$. Barring rapid oscillation
	of these quantities	as functions of $\phi$, 
	or a narrow spike,
	we certainly need them to be much less than unity to preserve $|\eta|\ll 1$
	over many e-folds. At a generic point it is reasonable to
	expect \cite{Kohri:2007qn} $|\sigma^3| \ll |\xi^2| \ll |\eta|$, but that
	will obviously fail if $\eta$ goes through zero.

 	Using~\eref{eq:zeta} to compute $\langle \zeta \zeta \zeta \rangle$
	and using Eq.~\eref{eq:fnl-def} for $\fnl$ gives
	\begin{eqnarray}
		\fl\nonumber
		\frac{6}{5} \fnl = \frac{3}{2} \epsilon_\ast - \eta_\ast +
		\xi_\ast^2 \left( \frac{1}{3} + \gamma + N_\ast \right) \\ \mbox{} +
		\frac{\epsilon_\ast}{\sum_i k_i^3} \left\{
			3 \frac{\xi_\ast^2}{\epsilon_\ast} \left( k_t \sum_{i < j} k_i k_j -
				\frac{4}{9} k_t^3 \right)
			+ \frac{4}{k_t} \sum_{i < j} k_i^2 k_j^2 +
			\frac{1}{2} \sum_{i \neq j} k_i k_j^2
		\right\} ,
		\label{eq:fnl}
	\end{eqnarray}
	where we recall that $N_\ast$ measures the number of e-folds which
	elapse between horizon exit of the mode $k$ and the time of evaluation,
	$\eta_\ast$.
	
	If taken at face value, Eq.~\eref{eq:fnl} may suggest that we can obtain
	an $\fnl$ as large as we please by allowing a large number of
	e-foldings outside the horizon, making $N_\ast$ very large. However, 
	that cannot be the case because the separate universe approach shows that
	$\zeta$ is conserved after horizon exit during single-field inflation.
	(More generally, $\zeta$ is conserved outside the horizon
	if isocurvature perturbations are negligible so that there is
 	a unique relation between pressure and energy density
	\cite{Wands:2000dp,Lyth:2004gb,Rigopoulos:2003ak}.) 
	From \eref{deta} we 
	see that the explicit $N_\ast$ dependence of $\fnl$ is cancelled by a
	contribution to the $N_\ast$ dependence of $\eta_\ast$. The remaining time
	dependence of $\fnl$ involves contributions proportional to
	$\epsilon_\ast\eta_\ast$, $\eta_\ast^2$ and
	$\sigma_\ast^3$, which would be cancelled if we took the calculation to a
	higher order in the field perturbation and abandoned the slow-roll
	approximation.

	The present example is trivial because $\zeta$ is time-independent.
	However, a similar caution applies to multiple-field inflation
	models where $\zeta$ is time-dependent. There also,
	powers of $N_\ast$ may also appear in the expression
	for $\fnl$. If these terms are to source large non-gaussianity
	(which occurs, for example, in the curvaton scenario
	\cite{Lyth:2002my,Lyth:2006gd,Malik:2006pm}), then they should
	be handled using a non-perturbative approach rather than relying
	on ``na\"{\i}ve'' expressions such as Eq.~\eref{eq:fnl}.
	
	\section{Conclusions}
	\label{sec:conclude}
	
	In this paper we have shown that the bispectrum of an interacting
	but canonically normalized scalar field can be calculated during an
	inflationary epoch using the details of its field equation directly,
	without constructing an effective action and using the rules of the
	in--in formalism.
	
	Where a Lagrangian formulation exists, the action and the field equations
	to which it gives rise are manifestly equivalent.
	In this case, as a point of principle, our calculation is a straightforward
	rewriting of the usual one, although it may prove to be more convenient
	in examples where one works with the field equation from the outset.
	On the other hand, in models where no Langrangian
	formuation exists, or none is known, our formula enables the
	non-gaussianity to be computed easily and compared with experiment.
	
	The recipe is straightforward. For reference, we summarise the
	steps here.
	\begin{itemize}
		\item Write the model in terms of a scalar field $\phi$, and
		determine the field equation for the perturbation in this field,
		$\delta\phi$, defined in the uniform curvature gauge.
		\item Separate $\delta\phi$ into a term $\delta\phi_1$ which obeys
		a linear equation of motion, and a part $\delta\phi_2$ which is
		quadratic in $\delta\phi_1$. Write down the evolution equation for
		$\delta\phi_2$.
		\item Determine the Fourier coefficients $\F_2$ and $\G_2$
		(respectively) of the terms quadratic in $\delta\phi_1$ and
		$\delta\phi_1'$.
		\item Construct the quantities $\symF_2$ and $\symG_2$
		from $\F_2$ and $\G_2$.
		\item Insert $\symF_2$ and $\symG_2$ in
		Eqs.~\eref{eq:f-correlator} and~\eref{eq:g-correlator},
		respectively. Simplify the resulting expressions to give the
		complete three-point expectation value.
	\end{itemize}
	
	We have stressed that the computation of the $\delta\phi$ expectation
	value soon after horizon crossing is a simple exercise.
	However, we wish to emphasize that by itself this is insufficient
	to obtain a prediction for the non-gaussianity which is observed in the
	CMB. For that, one needs to make a prediction for the expectation
	values of the curvature perturbation, $\zeta$, using the
	the classical evolution on superhorizon scales and the $\delta\phi$
	expectation values evaluated soon after horizon crossing as an initial
	condition. This process avoids the appearance of large logarithms
	which would require resummation.
	
	In the context of Einstein gravity coupled to a single scalar field
	this is simple to implement. The separate universe picture can be used
	in conjunction with rigorous results concerning the conservation of
	$\zeta$ on superhorizon scales to make robust predictions, especially
	in the case where isocurvature modes are absent. In more general
	theories it may be necessary to augment this procedure or find a
	replacement in order to make reliable estimates of the non-gaussianity
	which can actually be observed in the CMB.
	
	\ack
	
	We would like to thank Jim Lidsey for useful discussions.
	DS would like to thank Misao Sasaki and the Yukawa Institute of
	Theoretical Physics, Kyoto,
	for their hospitality during the September 2007
	programme \emph{Gravity and Cosmology}, where part of this work
	was completed. DS is grateful to the Astronomy Unit at
	Queen Mary, University of London, for continued hospitality.
	We would like to thank Neil Barnaby, Raghavan Rangarajan
	and Wei Xue for pointing out some errors and typos in the
	original version of this paper.
		
	\section*{References}
		

\providecommand{\href}[2]{#2}\begingroup\raggedright\endgroup

\end{document}